\documentclass[12pt]{article}
\usepackage{epsfig, amsmath, amssymb}
\usepackage{graphicx,psfrag} 
\newcommand{\be}{\begin{equation}}
\newcommand{\ee}{\end{equation}}
\newcommand{\ben}{\begin{equation}}
\newcommand{\een}{\end{equation}}
\newcommand{\bea}{\begin{eqnarray}}
\newcommand{\eea}{\end{eqnarray}}
\newcommand{\bA}{\begin{array}}
\newcommand{\eA}{\end{array}}
\newcommand{\bc}{\begin{center}}
\newcommand{\ec}{\end{center}}
\newcommand{\al}{\alpha}

\newcommand{\ra}{\rightarrow}
\newcommand{\del}{\partial}

\newcommand{\ie}{{\it i.e.}}
\newcommand{\eg}{{\it e.g.}}

\newcommand{\ua}{\uparrow}
\newcommand{\da}{\downarrow}
\newcommand{\lan}{\langle}
\newcommand{\ran}{\rangle}
\newcommand{\ura}{\underrightarrow}
\newcommand{\ula}{\underleftarrow}

\numberwithin{equation}{section}

\begin{document}

%\ifeprint
%\fi

\begin{titlepage}
%\vspace{30mm}

\bc

%%\hfill  {TIFR/TH/09-12} \\
\hfill % {\tt arXiv:0909.4731 [hep-th]} 
\\         [30mm]
%%X\vfill

{\Huge On extremal surfaces and de Sitter entropy}
% \\ [2mm] and  } 
\vspace{16mm}

{\large K.~Narayan} \\
\vspace{3mm}
{\small \it Chennai Mathematical Institute, \\}
{\small \it SIPCOT IT Park, Siruseri 603103, India.\\}
%%{\small Email: \ narayan@cmi.ac.in}\\

\ec
%\medskip
\vspace{40mm}

\begin{abstract}
We study extremal surfaces in the static patch coordinatization of de
Sitter space, focussing on the future and past universes. We find
connected timelike codim-2 surfaces on a boundary Euclidean time slice
stretching from the future boundary $I^+$ to the past boundary
$I^-$. In a limit, these surfaces pass through the bifurcation region
and have minimal area with a divergent piece alone, whose coefficient
is de Sitter entropy in 4-dimensions. These are reminiscent of rotated
versions of certain surfaces in the $AdS$ black hole. We close with
some speculations on a possible $dS/CFT$ interpretation of 4-dim de
Sitter space as dual to two copies of ghost-CFTs in an entangled
state. For a simple toy model of two copies of ghost-spin chains, we
argue that similar entangled states always have positive norm and
positive entanglement.
\end{abstract}

\end{titlepage}

%\newpage 
%{\tiny %footnotesize
%\begin{tableofcontents}
%\end{tableofcontents}
%}

%\vspace{5mm}

\section{Introduction}

de Sitter space is fascinatingly known to have temperature and
entropy \cite{Gibbons:1977mu}: see \eg\ \cite{Spradlin:2001pw} for a
review. This is most easily seen in the static patch coordinatization
of de Sitter space $dS_{d+1}$,
\be\label{dSstatic}
ds^2 = -\Big(1-{r^2\over l^2}\Big) dt^2 + {dr^2\over 1-{r^2\over l^2}}
+ r^2 d\Omega_{d-1}^2\ ,
\ee
as we review in sec.~\ref{dSentRev}.
\begin{figure}[h] 
\hspace{4pc} \includegraphics[width=4pc]{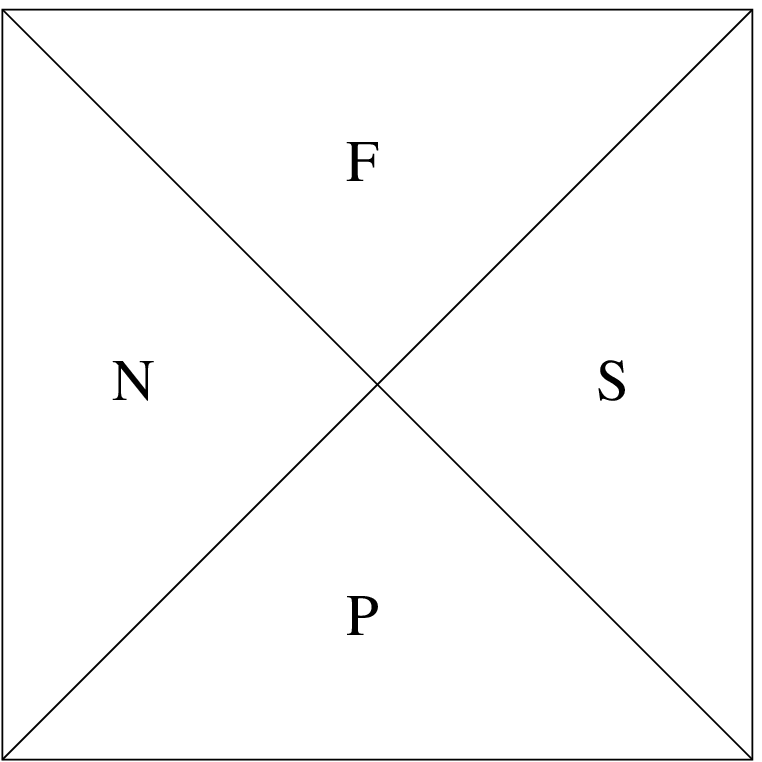} \hspace{3pc}
\begin{minipage}[b]{25pc}
  \caption{{\label{dSstat}\footnotesize {Penrose diagram of de Sitter space
        in static coordinates:  \newline  $N$ and $S$ are the Northern and
        Southern hemispheres. \newline $F$ and $P$ are future and past
        universes.  %\newline
}}}
\end{minipage}
\end{figure}
The Penrose diagram in Figure~\ref{dSstat} represents the
$(t,r)$-plane: each point is an $S^{d-1}$. The static patches refer to
the Northern and Southern hemisphere regions $N$ and $S$ where $t$ is
a timelike coordinate and time translations are Killing
isometries. $N$ and $S$ have $0\leq r \leq l$, with $r=0$ the Poles
and $r=l$ the cosmological horizon.  In the future and past universes
$F$ and $P$, $t$ is a spacelike direction and $r$ becomes time so
constant $r$ surfaces are spacelike.

de Sitter entropy is essentially the area of the cosmological horizon,
apparently stemming from degrees of freedom not accessible to
observers in regions $N$ and $S$, for whom the horizons are event
horizons. It is of interest to understand if this entropy can be
realized in gauge/gravity duality
\cite{Maldacena:1997re,Gubser:1998bc,Witten:1998qj,Aharony:1999ti}
for de Sitter space or $dS/CFT$
\cite{Strominger:2001pn,Witten:2001kn,Maldacena:2002vr} which
associates de Sitter space to a dual nonunitary Euclidean CFT on the
future (or past) boundary. In particular one might wonder if de Sitter
entropy is encoded in some generalization of the holographic
formulation of entanglement entropy
\cite{Ryu:2006bv,Ryu:2006ef,HRT,Rangamani:2016dms} via the areas of
appropriate extremal surfaces.

In ordinary (static) quantum systems, we consider spatial subsystems
on a constant time slice and entanglement entropy then arises by a
partial trace over the environment. Towards mimicking this in the dual
Euclidean CFT here, extremal surfaces on a constant boundary Euclidean
time slice were studied in \cite{Narayan:2015vda,Narayan:2015oka}:
these are anchored at the future boundary $I^+$ in the upper Poincare
patch of de Sitter space and dip into the bulk time direction. Real
extremal surfaces are either null with (minimal) vanishing area or
timelike with maximal area stemming from a divergent term alone.
However the areas of certain complex codim-2 extremal surfaces
(involving an imaginary bulk time parametrization) were found to have
structural resemblance with entanglement entropy in the dual Euclidean
CFT for the upper patch (defined in terms of some boundary Euclidean
time direction): in $dS_4$, these areas are negative, consistent with
the negative central charge \cite{Maldacena:2002vr} in
$dS_4/CFT_3$. These end up being equivalent to analytic continuation
from the Ryu-Takayanagi expressions in $AdS/CFT$. Further
investigations \cite{Narayan:2016xwq,Jatkar:2016lzq,Jatkar:2017jwz}
study generalizations of entanglement entropy for theories with
negative norm states, in particular ghost systems.

The present static patch coordinatization of de Sitter space,
Figure~\ref{dSstat}, raises the possibility of de Sitter entropy
arising from the area of appropriate connected extremal surfaces
stretching from the boundary of some subregion at the future boundary
$I^+$ of $F$ to some equivalent subregion at the past boundary $I^-$
of $P$.  Naively this might then allow an interpretation of the
surface as representing some ``generalized entanglement'' between the
future and past copies of the Euclidean CFT. These are reminiscent of
certain surfaces studied by Hartman and Maldacena
\cite{Hartman:2013qma} in the $AdS$ black hole, except the present
surfaces stretching in the time direction between $I^+$ and $I^-$ are
in some sense rotated versions thereof\ (similar surfaces were in fact
studied in \cite{Narayan:2015vda} in the de Sitter bluewall
\cite{Das:2013mfa} as we discuss below).

We review the static patch and de Sitter entropy in sec.~2 and a
version of entanglement as interface area.  In sec.~3, we describe
extremal surfaces in $F$ and $P$: we find that 4-dim de Sitter entropy
${l^2V_{S^2}\over 4G_4}={\pi l^2\over G_4}$ arises as the coefficient
of the divergent area of certain connected codim-2 real timelike
extremal surfaces lying in a boundary Euclidean time slice of the bulk
space, stretching from $I^+$ to $I^-$. These pass through the
bifurcation region and have minimal area. It is worth noting that the
horizons from the point of view of $F$ and $P$ are not event horizons
but Cauchy horizons, so that this recovery of de Sitter entropy might
appear unconventional. We close with some speculations (sec.~4) on a
$dS/CFT$ interpretation of $dS_4$ as dual to two copies of ghost-CFTs
on $I^+$ and $I^-$ in particular entangled states. Appendices A,B
review aspects of Poincare $dS$ extremal surfaces and ghost-spins.

\section{The static patch and de Sitter entropy}\label{dSentRev}

The Euclidean continuation $t\ra -it_E$ of (\ref{dSstatic}) gives\
$ds_E^2 = {dr^2\over 1-r^2/l^2} +(1-{r^2\over l^2}) dt_E^2 
+ r^2 d\Omega_{d-1}^2$\ which is a sphere, and de Sitter entropy can be
obtained as for black holes. Regularity requires that there be no
conical singularity in the $(t,r)$-plane at the origin (which was the
location of the horizon). This makes $t_E$ an angular variable with
periodicity $2\pi l$ which is the inverse Hawking temperature of de Sitter
space. With the horizon as one boundary, the Euclidean action gives
\be
I_E = - \int {r^{d-1} dr dt_E d\Omega_{d-1} \over 16\pi G_{d+1}}
\left(R-{d(d-1)\over l^2}\right)\
%= - \int {dr dt_E d\Omega_{d-1} r^{d-1}\over 16\pi G_{d+1}}\ {2d\over l^2}
%\nonumber\\
=\ -{V_{S^{d-1}} (2\pi l)\over 16\pi G_{d+1}} {2d\over l^2} {r^d\over d}\Big|_0^l
\ =\ -{l^{d-1} V_{S^{d-1}}\over 4G_{d+1}}\ .
\ee
Equivalently, the Euclidean continuation of $dS_{d+1}$ is $S^{d+1}$
with $ds^2=l^2ds_{S^{d+1}}^2$: this gives\
$I_E = -\int { d\Omega_{d+1} l^{d+1}\over 16\pi G_{d+1}} {2d\over l^2}$\
using\ $V_{S^d}={2\pi^{(d+1)/2}\over\Gamma((d+1)/2)}$\
and\ $V_{S^{d+1}} = {2\pi\over d} V_{S^{d-1}}$.\ 
Since the sphere has no boundary, there is no ``energy'' contribution
to $I_E$ so $E=0$, giving
\be\label{dSent}
\log Z = -\beta F = -\beta E + S = -I_E \qquad \Rightarrow\qquad
S_{dS_{d+1}} = -I_E = {l^{d-1} V_{S^{d-1}}\over 4G_{d+1}}\ ,
\ee
giving the entropy of de Sitter space. For $dS_4$, this is\
$S_{dS_4} = {\pi l^2\over G_4}$~.

%S_{dS_3} = {\pi l\over 2G_3}\ ,\qquad S_{dS_5} = {\pi^2l^3\over 2G_5}\ .

%\noindent {\bf Static patch and entanglement entropy as interface area:}\ \

Since the regions $N$ and $S$ are static with $t$-translations being
isometries, it is natural to ask if there are extremal surfaces which
wrap the horizon in the IR limit and whose area recovers de Sitter
entropy. We recall that in the $AdS$ black brane, the Ryu-Takayanagi
minimal surface \cite{Ryu:2006bv} wraps the horizon in the IR limit
where the subsystem approaches the full space and the finite part of
entanglement entropy given by the minimal surface area approaches the
entropy of the black brane given by the horizon area. Thus consider a
constant time slice $t=const$: the spatial metric is 
\be
d\sigma^2 = {dr^2\over 1-r^2/l^2} + r^2 d\Omega_{d-1}^2
= l^2 ( d\theta^2 + \sin^2\theta d\Omega_{d-1}^2 )\ ,
\ee
where the Southern\ hemisphere has\ $r=l\sin\theta$\ with
$0\leq r\leq l$ so $0\leq \theta\leq {\pi\over 2}$ and likewise for
the Northern\ hemisphere.
%r=-l \sin\theta\ , \qquad \qquad -l\leq r\leq 0 \quad \Rightarrow\quad
%0\leq \theta\leq {\pi\over 2} \ .
Thus the $t=const$ slice of the static patch is a sphere comprising
two hemispheres with $r$ the latitudinal coordinate. The horizons at
$r^2=1$ are then the equators at the boundary $\theta={\pi\over 2}$
of the hemisphere (with the Poles at $r=0$ or $\theta=0$). Notably,
unlike $AdS$, the coordinate $r$ is not a bulk radial coordinate but
simply an angular direction on the hemisphere (which has no boundary):
latitudes at $r=const$ define hemispherical caps which might be the
most natural subsystems here.

Each latitude divides the full sphere into the subsystem and the rest
of the sphere (which comprises the rest of the hemisphere containing
the cap and the other hemisphere). It is intuitive to define the
entanglement entropy as the interface area in Planck units.  The
latitude defined by the equator at $r=l$ divides the full sphere into
a subsystem defined by one hemisphere and the environment which is the
other hemisphere. Then the entanglement entropy between the two
hemispheres becomes de Sitter entropy,
\be
S_r = {r^{d-1} V_{S^{d-1}}\over 4G_{d+1}}\quad\ra\quad
{l^{d-1} V_{S^{d-1}}\over 4G_{d+1}}\ .
\ee
In this limit, the interface is in fact the horizon, and so this
agrees with the familiar statement that de Sitter entropy is the area
of the cosmological horizon in Planck units. See also \eg\
\cite{Hawking:2000da}, \cite{Nguyen:2017ggc}.

\section{The future and past universes and extremal surfaces}

The future and past de Sitter universes in (\ref{dSstatic}) with\
$1\leq {r\over l}\leq \infty$\ can be described as
\be\label{dSst}
ds^2 = {l^2\over\tau^2} \left(-{d\tau^2\over 1-\tau^2} + (1-\tau^2) dw^2
+ d\Omega_{d-1}^2\right) ,    %0\leq \tau\leq 1 ,
\qquad\qquad \ \tau={l\over r}\ ,\quad  w={t\over l}\ .
\ee
$\tau$ is now the ``bulk'' time coordinate while $w$ is a spatial
coordinate enjoying translation invariance. The maximal extension
encoded in the Penrose diagram exhibits horizons at $\tau=1$, which
are Cauchy horizons for the future and past universes $F$ and $P$. For
instance there are trajectories which end in $N$ or $S$ so they cannot
be part of the Cauchy data whose time development leads to $I^+$: thus
the future horizons are past Cauchy horizons for data on $I^+$ acting
as causal boundaries for the future universe $F$ cloaking the static
patches.  Likewise the past horizons are future Cauchy horizons for
Cauchy data on $I^-$\ (related discussions appear in the de Sitter
bluewall \cite{Das:2013mfa}). The future and past boundaries $I^+$
and $I^-$ are at $\tau=0$. The asymptotic structure of the future
universe is
\be
{r\over l}\gg 1:\qquad\qquad
ds^2 \sim\ - l^2{dr^2\over r^2} + {r^2\over l^2} dt^2 + r^2 d\Omega_{d-1}^2\
\sim\  {l^2\over\tau^2} \left (-d\tau^2 + dw^2 + d\Omega_{d-1}^2 \right)\ . 
\ee
This is akin to the Poincare patch of de Sitter space except that the
boundary is not $R^d$ but Euclidean $R\times S^{d-1}$. This is analogous
to global $AdS_{d+1}$ where the boundary is $R_{time}\times S^{d-1}$.
Constant $\tau$ slices (\ie\ $r=const$ slices in (\ref{dSstatic}))
have topology $R\times S^{d-1}$.\

%\subsection{Extremal surfaces}

We want to look for extremal surfaces stretching from $I^+$ to $I^-$
whose area might capture de Sitter entropy.
The scaling ${l^{d-1}\over G_{d+1}}$ of de Sitter entropy
suggests that the surfaces in question are codimension-2.  From the
point of view of entanglement in the dual theory defined with respect
to Euclidean time, it would seem reasonable to look for bulk surfaces
lying on an appropriately defined constant boundary Euclidean time
slice of the bulk space.  Noting that the space enjoys $t$-translation
symmetry as well as rotational invariance in $S^{d-1}$, let us
imagine restricting to (i) an equatorial plane of the $S^{d-1}$, or
(ii) a $t=const$ surface as a constant boundary Euclidean time
slice. \vspace{1mm}

%\noindent {\underline {\bf An $S^{d-1}$ equatorial plane}:\ \

\subsection{An $S^{d-1}$ equatorial plane}

\begin{figure}[h] 
\hspace{0.1pc}
\includegraphics[width=15pc]{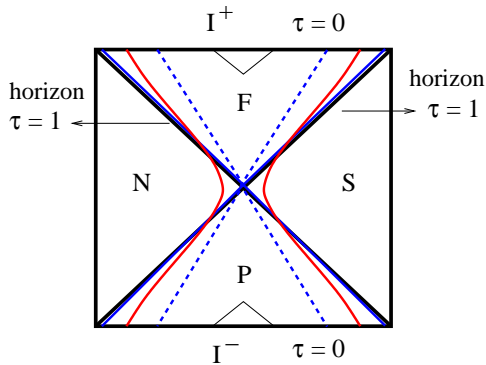} \hspace{0.5pc}
\begin{minipage}[b]{30pc}
  \caption{{\label{dSstSurfEnt}\footnotesize {Timelike extremal surfaces
        in the $(\tau,w)$-plane \newline stretching from  $I^+$ to
        $I^-$. These are akin to rotated versions \newline of
        the surfaces of Hartman-Maldacena in the $AdS$ black hole.
        \newline The red timelike surface intersects the horizons.
        \newline The limiting blue surface is almost null (almost hugging
        \newline the horizons) and passes through the bifurcation
        region: \newline it has minimal area.
        \newline
}}}
\end{minipage}
\end{figure}

We restrict
to an equatorial plane with $\theta={\pi\over 2}$. The rotational symmetry
implies that all such equatorial planes are equivalent. The metric on
such a slice from (\ref{dSst}) is
\be\label{eqSlicedS}
ds^2 = - {l^2\over\tau^2} \left({d\tau^2\over 1-\tau^2} + (1-\tau^2) dw^2
+ d\Omega_{d-2}^2\right)\ .
\ee
This equatorial slice can be thought of as follows: the future component
comprises a family of concentric cylinders $R_w\times S^{d-2}$ at
$\tau=const$ slices, with size ${l\over\tau}$ and $\tau$ representing
the radial direction, the outermost cylinder having size
${l\over\epsilon}$ while the innermost has size $l$. The past component
comprises a similar family with $\tau$ again running over
$l\leq\tau\leq {l\over\epsilon}$~. The two join at the bifurcation
region (the intersection of the horizons) with $\tau=1$: this is a
smooth $S^{d-2}$ as can be seen via Kruskal-type coordinates with $y$
the ``tortoise'' coordinate 
\be\label{tortoise}
u=e^{w-y} ,\quad v=-e^{-w-y} ,\qquad\
y=\int {d\tau\over 1-\tau^2} = {1\over 2} \log \Big|{1+\tau\over 1-\tau}\Big|
\qquad\qquad [0<\tau<\infty]\ .
%\quad [0\leq\tau< 1]\ ;\qquad y = {1\over 2} \log {\tau+1\over \tau-1}\quad
\ee
We want to impose boundary conditions that reflect extremal surfaces
stretching from the boundary of a subsystem of the form
$\Delta w\times S^{d-2}$ at $I^+$ dipping into the bulk, to finally end
at the boundary of an equivalent subsystem at $I^-$, as in
Figure~\ref{dSstSurfEnt}.
These could either penetrate the horizons somewhere, or pass through the
bifurcation region without intersecting the horizons (as do all static
observers at any fixed $w$). With ${dw\over d\tau}\equiv w'$, the area
functional is
\be\label{areaFnEquator}
%= \int {l^{d-1}V_{S^{d-2}}\over \tau^{d-1}} \sqrt{{d\tau^2\over 1-\tau^2} -
%(1-\tau^2) dw^2}\
S=\ l^{d-1} V_{S^{d-2}} \int {d\tau\over\tau^{d-1}}
   \sqrt{{1\over 1-\tau^2} - (1-\tau^2) (w')^2}\ .
\ee
The subsystems in question are on $I^+$/$I^-$ and so spacelike: we
therefore take these to be real surfaces orthogonal to the subsystems
and so timelike at least initially (\ie\ $w'\sim 0$ near the boundary
$\tau=0$ and $S\sim \int l^{d-1} {d\tau\over\tau^{d-1}}$), thereby
choosing the sign under the square root.

\noindent Along the lines of the de Sitter bluewall 
\cite{Das:2013mfa} analysed in \cite{Narayan:2015vda}, we obtain
\be
   {-(1-\tau^2) w'\over \sqrt{{1\over 1-\tau^2} - (1-\tau^2) (w')^2}}
   {1\over \tau^{d-1}} = B
   \quad \Rightarrow \quad
{1\over 1-\tau^2} - (1-\tau^2) (w')^2 = {1\over 1-\tau^2 + B^2\tau^{2d-2}}\ ,
\ee
giving\ ($I^\pm$ are regulated at $\epsilon$, and $\tau_*$ is the
turning point discussed below)
\be\label{HMsurf}
{\dot w}^2 \equiv (1-\tau^2)^2 (w')^2 = {B^2\tau^{2d-2}\over 1-\tau^2
           + B^2\tau^{2d-2}}\ ,
\qquad S = 2 l^{d-1} V_{S^{d-2}} \int_\epsilon^{\tau_*} {d\tau\over\tau^{d-1}}\
        {1\over \sqrt{1-\tau^2 + B^2\tau^{2d-2}}}\ .
 \ee
Here $B$ is a conserved constant.
The factor of $2$ in the area arises because we are calculating the areas
for both the top and bottom half-surfaces (see Figure~\ref{dSstSurfEnt}).
The structure of these extremal surfaces is somewhat different from 
those in the Poincare slicing \cite{Narayan:2015vda} (reviewed briefly
in Appendix~\ref{sec:extSurPoinc}): for sufficiently small subsystems
however, (\ref{HMsurf}) approaches (\ref{RTdS02}).

In (\ref{HMsurf}),\ ${\dot w}=(1-\tau^2){dw\over d\tau}$ is the
$y$-derivative with $y$ in
(\ref{tortoise}) above, useful in the vicinity of the horizons.
Near the boundary $\tau\ra 0$, we have ${\dot w}\ra 0$ for any nonzero
finite $B$: further we have ${\dot w}<1$ for $\tau<1$
(within the future universe), \ie\ the surface drops down towards the
past.  As $\tau\ra 1$, we have ${\dot w}\ra 1$,
\ie\ the surface grazes the horizon (is tangent to the horizon) when it
intersects the horizon.  $\tau>1$ gives ${\dot w}>1$: this is
in the regions $N$ or $S$, after crossing the horizon.\ Note that
$\tau=const$ surfaces within $N$ or $S$ are timelike.

The turning point $\tau_*$ is the ``deepest'' location to which the
surface dips into in the bulk, before turning around: this is
when\ ${\dot w}\ra\infty$ or ${dy\over dw}=0$. Thus the surface has no
variation in the $\tau$- (or $y$-)direction, \ie\ it is tangent to
$\tau=const$ curves at the location $\tau_*$\ (see the red curve in
Figure~\ref{dSstSurfEnt}).  From (\ref{HMsurf}), we see that for
$\tau<1$ (within $F$ or $P$), the surface has no tendency to turn
around since ${\dot w}<1$: however the surface can have a turning
point if $\tau>1$.  To see this explicitly, we note using
(\ref{tortoise}) in (\ref{HMsurf}) that
\be\label{HMsurfy}
\tau>1:\qquad\qquad {\dot w}^2 = \Big({dw\over dy}\Big)^2
= 1\Big/\Big(1 - {4e^{2y} (e^{2y}-1)^{2d-4}\over B^2 (e^{2y}+1)^{2d-2}}\Big)\ .
\ee
$y\ra\infty$ at the $\tau=1$ horizons and ${\dot w}^2\ra 1$\ \ie\
$w\ra \pm y$\ (the area functional (\ref{areaFnEquator}) recast using
the Kruskal coordinates $u,v$ in (\ref{tortoise}) can be seen
to be regular at the $\tau=1={1-uv\over 1+uv}$ horizons, \ie\ $u=0$
or $v=0$).\ The turning point is at
\be
 |{\dot w}|\ra\infty:\qquad\qquad
 1-\tau_*^2+B^2\tau_*^{2d-2}\ =\ 0\ =\
 1 - {4e^{2y_*} (e^{2y_*}-1)^{2d-4}\over B^2 (e^{2y_*}+1)^{2d-2}}\ .
\ee
At $\tau_*$, the surface from $I^+$ is joined with the surface from
$I^-$. We have\ $w\sim\sqrt{\tau_*-\tau}$\ from above (near
$\tau\lesssim\tau_*$) joining $w\sim -\sqrt{\tau_*-\tau}$\ from below
smoothly.  This gives the full smooth ``hourglass''-shaped surfaces
in Figure~\ref{dSstSurfEnt}.  Thus the surface
starts at $w=\pm {\Delta w\over 2}$ on $I^+$, intersects the future
horizons at $\tau=1, w=\pm\infty$, turns around at $\tau_*$ in
$N$/$S$, then intersects the past horizons at $\tau=1, w=\mp\infty$,
finally reaching $w=\mp {\Delta w\over 2}$ on $I^-$.
 
We will now mostly focus on 4-dim de Sitter space $dS_4$
(\ie\ $d=3$) which turns out to be most interesting. From
(\ref{HMsurf}), since\ $1-\tau^2+B^2\tau^4=(1-B\tau^2)^2+(2B-1)\tau^2$,\
we have
\be\label{tau*B}
0<B<{1\over 2}:\qquad\qquad
{\dot w}\ra\infty\ ,\qquad\qquad 1-\tau_*^2+B^2\tau_*^4=0\ ,
\ee
giving the turning point $\tau_*(B^2)$ as a function of the
parameter $B^2$.
For $B=0$, we have $\tau_*=1$: small $B$ gives $\tau_*\gtrsim 1$.
The limit $B\ra 0$ gives $\tau_*\ra 1$, \ie\ the surface
has a turning point in the interior very close to the bifurcation
region (the red ``hourglass neck'' is pinching off), at
\be\label{smallB}
\tau_*=1+\delta\qquad \Rightarrow\qquad -2\delta + B^2 \sim 0 \sim
1 - {4\over B^2e^{2y_*}} \quad
\Rightarrow \quad \delta \sim {B^2\over 2}\ .
\ee
So $\delta\ra 0$ as $B\sim 0$, with ${\dot w}\sim 0$.\
From (\ref{HMsurf}), the area for $B=0$ in Planck units becomes
\be\label{EEdSent}
   {S\over 4G_{d+1}} = {2 l^{d-1} V_{S^{d-2}}\over 4G_{d+1}}
   \int_\epsilon^1 {d\tau\over \tau^{d-1}}\ {1\over \sqrt{1-\tau^2}}\ .
\ee
The surface stretches from the boundary $\tau=\epsilon$ in the future
universe to the bifurcation region with $\tau_*=1$ and then has a similar
piece in the past universe. For $dS_4$, the surface is a 1-dimensional
curve in the $(\tau,w)$-plane and wraps the $S^1$ on a $\tau=const$
slice. This gives
\be\label{EEdS4ent}
{S\over 4G_4}\ =\
{2l^2 (2\pi)\over 4G_4} \left({-\sqrt{1-\tau^2}\over\tau}\right)\Big|_\epsilon^1
%=\ {\pi l^2\over G_4} {1\over\epsilon}\
\ =\ {\pi l^2\over G_4} {l\over\epsilon_c}\ .\qquad\qquad[\epsilon_c=l\epsilon]
\ee
The coefficient of this divergent area in Planck units is
precisely de Sitter entropy. The bulk surface stretches from the
future boundary $I^+$ regulated at $\epsilon={\epsilon_c\over l}$
(expressed in terms of the de Sitter scale $l$), and passes 
through the bifurcation region at $\tau=1$. The turning point is
contained in the bifurcation region, and so is fixed at $\tau_*=1$.
Interpreting this as an area law divergence in a dual CFT however,
rescaling the ultraviolet cutoff changes the precise coefficient.
%\ie\ $\tau\ra \lambda \tau$ gives $\tau_*=1\ra \tau_*=\lambda$ and
%$\epsilon\ra\lambda\epsilon$,\   %\ie\ $r_*=l$.

It is also interesting to study the width of the subregion as $B\ra 0$
(I thank Veronika Hubeny for a discussion which led to this). From
(\ref{tau*B}), (\ref{tortoise}),
\be
B^2 = {\tau_*^2-1\over\tau_*^4} = {4e^{2y_*} (e^{2y_*}-1)^2\over (e^{2y_*}+1)^4}
\ee
so as the turning point $\tau_*\ra 1$, we have $y_*\ra \infty$, with
$B^2\sim 4e^{-2y_*}$.\ Thus from (\ref{HMsurf}), (\ref{HMsurfy}), in the
limit $B\ra 0$, we see that the width $\Delta w(y_*)$ scales as
\be
\Delta w = 2 \int_0^{y_*}
dy\Big/\sqrt{1 - {4e^{2y} (e^{2y}-1)^{2d-4}\over B^2 (e^{2y}+1)^{2d-2}}}\ 
\sim\ 2\int^{y_*} {dy\over\sqrt{1-e^{2y_*}/e^{2y}}}\ \sim\ \log e^{2y_*} \sim
2y_*\ ,
\ee
from the contribution near the turning point $y\sim y_*$ (which is large).
In other words, in the limit $B\ra 0$, we have
$\Delta w\sim 2y_*\sim \log{2\over\tau_*-1}\sim 2\log{2\over B}\ra\infty$,
\ie\ the subregion, defined by the boundaries of the surface, becomes
all of $I^{\pm}$ (on the equatorial plane). The area (\ref{HMsurf}),
written as\ $\int_\epsilon^1(\ldots)+\int_1^{\tau_*}(\ldots)$ arises
mostly from the first term giving (\ref{EEdSent}), the second term
(with near vanishing area) corresponding to the limiting surface
almost tracing the horizon.

\subsection{Features of $S^{d-1}$ equatorial plane extremal surfaces}

In general, de Sitter space does not appear to exhibit interesting
solutions to extremization, unlike $AdS$: \eg\ in the Poincare slicing
(which is the local geometry near any point at $I^+$), as we have
mentioned earlier, surfaces do not have any real turning point,
reviewed briefly in Appendix~\ref{sec:extSurPoinc}\ (complex extremal
surfaces were found in \cite{Narayan:2015vda} which amount to analytic
continuation from Ryu-Takayanagi in $AdS$). The surfaces
(\ref{HMsurf}) circumvent this since they stretch from $I^+$ to $I^-$:
these surfaces are somewhat special, as we discuss below.

Firstly, it is interesting to note that the subsystem size does not
enter in (\ref{EEdSent}), (\ref{EEdS4ent}). In the limit $B\ra 0$, 
the subsystem becomes the entire space $\Delta w\ra\infty$: 
the smooth red curve in Figure~\ref{dSstSurfEnt} (\ie\ the
surface at generic $\Delta w$ or $B$) becomes the limiting blue curve
with $B\ra 0$ and $\Delta w\ra\infty$.  This surface has $\tau_*\ra 1$
and hugs the horizons without intersecting them: it just grazes the
future horizon dropping down from $I^+$, and then smoothly turns
around and hugs the past horizon to eventually hit $I^-$. The surface
thus appears to exclude precisely the regions behind the horizons,
\ie\ regions $N, S$, restricted to this equatorial plane.

These $B=0$ surfaces passing through the bifurcation region in fact
have minimal area. 
Firstly we mention that this is confirmed by numerical evaluation of
the area integral in (\ref{HMsurf}). Secondly, in the neighbourhood of
$B=0$, we can also analytically evaluate the change in the area: with
$S=\int_\epsilon^{\tau_*(B^2)} L(\tau,B^2)d\tau$, we have the first
order change for infinitesimal $\delta B^2$,
\be
\delta S = \Big(\int_\epsilon^{\tau_*(B^2)} {\del L\over \del B^2}(\tau_*(B^2))
d\tau + L(\tau_*(B^2),B^2) {\del\tau_*(B^2)\over\del B^2} \Big)\ \delta B^2\ .
\ee
For the $dS_4$ surfaces in (\ref{EEdSent}), (\ref{EEdS4ent})
with $B^2=0$, turning on a small $\delta B^2$ and using (\ref{smallB}),
we find that a term singular near $\tau_*=1$ cancels between both
terms in $\delta S$ giving\ $\delta S = {\pi\over 2} \delta B^2>0$.
Thus the deformation (\ref{smallB}) of the $B=0$ surface increases its area.

More broadly, for any $B>{1\over 2}$ we see that\
$1-\tau^2+B^2\tau^4=(1-B\tau^2)^2 + (2B - 1)\tau^2>0$\ does not
vanish: thus there is no turning point solution to (\ref{tau*B}).
The value $B={1\over 2}$ in $dS_4$ gives\ ${\dot w}^2 = {\tau^4/4\over
(1-\tau^2/2)^2}$~. Thus at this special value of $B$, the turning
point is at $\tau_*=\sqrt{2}$ where ${\dot w}\ra\infty$ with a double
zero in the denominator. The subregion width $\Delta w$ acquires a
divergence near $\tau_*=\sqrt{2}$: the surface area also has a
logarithmic divergence here.\ There is an accumulation of surfaces
with turning point near $\tau_*=\sqrt{2}$ which is a limiting value:
the surfaces appear to be ``repelled'' from dipping into the static
patches $N$ or $S$ to larger $\tau_*$ values.  For generic
$0<B<{1\over 2}$, the turning point is given by a single zero of the
polynomial in (\ref{tau*B}): of the two positive roots, we pick the
root satisfying\ $\tau_*<\sqrt{2}$ which is the limiting value. For
$B>{1\over 2}$ extremal surfaces stretching from $I^+$ to $I^-$ do not
exist: there are however \eg\ disconnected null surfaces with
$B\ra\infty$ lying entirely within $F$ (or $P$) as shown by the small
disconnected black wedges in Figure~\ref{dSstSurfEnt}.

\noindent Thus for any given subsystem
$(\Delta w\times S^1)^2 \in I^+\cup I^-$, we finally see that there are:
\begin{itemize}
  %$\bullet$\
{\item minimal (zero) area disconnected surfaces with $B\ra\infty$: from
(\ref{HMsurf}), these are null with ${\dot w}=1$ and vanishing area,
shown as the two disconnected black wedges in Figure~\ref{dSstSurfEnt}.
Each wedge is seen to have support only at one boundary ($I^+$ or $I^-$):
with $B$ large, the area in (\ref{HMsurf}) is\
$S\sim 2l^{d-1} V_{S^{d-2}} \int_\epsilon^{\tau_*} {d\tau\over\tau^{d-1}}\
{1\over \sqrt{1 + B^2\tau^{2d-2}}}$, and we see that there is
no smooth turning point. The black wedges are real null surfaces with
$S\ra 0$ as $B\ra\infty$ (half-surfaces joined with a cusp at $\tau_*$,
similar to the real null surfaces in \cite{Narayan:2015vda}).
Sufficiently small subregions can be approximated as akin to the flat
Poincare slicing, as is clear from the area approximation here, and so
also admit complex extremal surface solutions as in
\cite{Narayan:2015vda} with negative area.} %\\
%  $\bullet$\
{\item minimal area connected surfaces with finite $B$, and area
(\ref{EEdSent}), (\ref{EEdS4ent}) as $B\ra 0$: for generic $\Delta w$,
these are shown as the smooth red curve from $I^+$ to $I^-$ in
Figure~\ref{dSstSurfEnt}, with the limiting blue curve for $\Delta
w\ra\infty$ (as $B\ra 0$) cloaking the horizons. The surfaces with
infinitesimal $B$ in (\ref{smallB}) which intersect the horizons with
a smooth turning point just inside the horizon asymptote to this
$B=0$ surface.  From (\ref{HMsurf}), we see that this limiting surface
almost simply ``hangs down'' from $I^+$ till $I^-$ without bending or
turning since ${\dot w}\sim 0$.\ (The dotted blue lines are generic
curves with ${\dot w}=0$: the area is independent of the size of the
subregion for these surfaces, which pass through the bifurcation
region). The area in (\ref{HMsurf}) smoothly asymptotes to
(\ref{EEdSent}), (\ref{EEdS4ent}), as $B\ra 0$ and is then
minimal. These surfaces are reminiscent of the surfaces of
Hartman-Maldacena \cite{Hartman:2013qma} in the $AdS$ black hole:
perhaps this is not surprising since in some sense the de Sitter
static patch coordinatization is a rotation of the $AdS$ black hole
(although not an analytic continuation).}
\end{itemize}

\noindent Overall, for $B$ small, the term with the minus sign dominates
over the $B^2$-term under the square root so that the area integral in
(\ref{HMsurf}) (approaching (\ref{EEdSent})) bears resemblance to the
Ryu-Takayanagi area integral $S = 2R^{d-1}V_{d-2} \int_\epsilon^{r_*}
{dr\over r^{d-1}} {1\over\sqrt{1-r^{2d-2}/r_*^{2d-2}}}$ with $r_*\sim l$
for a strip subsystem of width $l$ in $AdS$: the full integral in fact
has some similarity to the area integrals for strip subsystems in the
$AdS$ plane wave geometry \cite{Narayan:2012ks}.  In some sense, the
neighbourhood of the bifurcation region in the static $dS$ patch
behaves like $AdS$ with regard to the area functional
extremization\footnote{While the Poincare slicing (Appendix A) must
  also exhibit these as geometric surfaces, they appear tricky to see
  directly.}. This singles out these minimal area connected surfaces
above in de Sitter space as special.

For other dimensions, the coefficient scales as $dS$ entropy but is not
precisely that:
\be
[dS_3]\ \
{S\over 4G_3} = {2.2l\over 4G_3}\int_\epsilon^1 {d\tau\over\tau}
{1\over\sqrt{1-\tau^2}} = {l\over G_3} \log {2\over\epsilon}\ ;\quad\ \
[dS_5]\ \ {S\over 4G_5}
= {\pi l^3\over G_5} \left({1\over\epsilon^2} + \log {2\over\epsilon} \right) .
\ee
This is not surprising: the surface in (\ref{EEdSent}) wraps an
$S^{d-2}$ and the horizon directions in the $(\tau,w)$-plane, which
compensates precisely for the $S^{d-1}$ in (\ref{dSent}) only for
$dS_4$ interestingly.

These surfaces bear some qualitative similarity to the trajectories of
timelike geodesics stretching from $I^+$ to $I^-$ in the
$(\tau,w)$-plane.  A $w=const$ timelike geodesic in the
$(\tau,w)$-plane is a straight line passing through the bifurcation
region. In general, timelike geodesics have action\ $S=\int
\sqrt{-g_{\tau\tau}d\tau^2 + g_{ww}dw^2}$~. Simplifying, we see that
this is identical to (\ref{areaFnEquator}) with $d=2$: thus the length
of such limiting geodesics is similar to the $dS_3$ result above.
%= l\int {d\tau\over\tau} \sqrt{{1\over 1-\tau^2} - (1-\tau^2)(w')^2}$.
Likewise for codim-1 surfaces, the area functional\
$S = l^{d}V_{S^{d-1}} \int {d\tau\over \tau^{d}}\
\sqrt{{1\over 1-\tau^2} - (1-\tau^2) (w')^2}$\ can be seen to scale as
$l^d$ in $dS_{d+1}$. Analysing this shows that codim-1 surfaces with
$B=0$ in $dS_4$ have\
$S = 2 l^{3} V_{S^{2}} \int_\epsilon^{1} {dr\over r^{3}} {1\over\sqrt{1-r^2}}
= 4\pi l^3 ({1\over\epsilon^2} + \log {2\over\epsilon})$.

\vspace{5mm}

%\vspace{10mm}

%\noindent {\underline {\bf The $w=const$ slice}}:\ \

\subsection{The $w=const$ slice}

From (\ref{dSst}), the metric on this slice becomes
\be
ds^2 = - l^2{d\tau^2\over \tau^2(1-\tau^2)} + {l^2\over\tau^2}
(d\theta^2+\sin^2\theta d\Omega_{d-2}^2)\ .
\ee
All $w=const$ slices pass through the bifurcation region.
The natural subsystem here is a cap-like region defined by a latitude
at $\theta=const$ on the $S^{d-1}$ at the future boundary.
\begin{figure}[h] 
\hspace{6pc} \includegraphics[width=6pc]{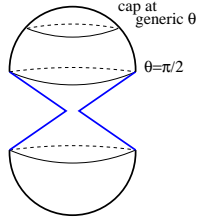} \hspace{3pc}
\begin{minipage}[b]{25pc}
  \caption{{\label{dSstSurfEnt2}\footnotesize {Timelike extremal surfaces
        in the $(\tau,\theta)$-plane \newline stretching from ${\cal I}^+$
        to ${\cal I}^-$. 
        The limiting blue surface \newline stretches from one hemisphere
        ($\theta={\pi\over 2}$) at $I^+$ to another \newline equivalent
        one at $I^-$.  \newline  %\newline
}}}
\end{minipage}
\end{figure}
The extremal surface we are looking for stretches from the boundary of
one such cap at $I^+$ to an equivalent cap on the $S^{d-1}$ at the past
boundary $I^-$, as in Figure~\ref{dSstSurfEnt2}.
Both caps wrap an $S^{d-2}$ on the $S^{d-1}$.
%It turns out that generic surfaces do not exhibit such behaviour but
%specific ones do, as we will describe.
The area functional is
\be\label{areaFnTheta}
S = 2 \int {l^{d-1} V_{S^{d-2}} (\sin\theta)^{d-2}\over \tau^{d-1}}
\sqrt{{d\tau^2\over 1-\tau^2} - d\theta^2}\
=\ 2l^{d-1} V_{S^{d-2}} \int {d\tau\over \tau^{d-1}} (\sin\theta)^{d-2}
\sqrt{{1\over 1-\tau^2} - (\theta')^2}\ .
\ee
The factor of 2 arises as before from the two components of the surface,
one stretching from $I^+$ and another from $I^-$.
The equation of motion\ ${d\over d\tau} ({\del L\over\del \theta'}) =
{\del L\over\del\theta}$\ becomes
\be\label{tconstEOM}
   {d\over d\tau} \left({-\theta'\over\sqrt{{1\over 1-\tau^2}-(\theta')^2}}
   {(\sin\theta)^{d-2}\over \tau^{d-1}}\right) =
   (d-2) {(\sin\theta)^{d-3}\over \tau^{d-1}}
   \cos\theta \sqrt{{1\over 1-\tau^2}-(\theta')^2}\ .
\ee
The analogs of the extremal surfaces earlier passing through the
bifurcation region with $w'=0$ in this case are surfaces which
``hang'' down into the bulk without turning, \ie\ with $\theta'=0$.
(By contrast, $\theta'$ maximum gives $\theta'={1\over\sqrt{1-\tau^2}}$,
which are null surfaces with vanishing (minimal) area.)
With $\theta'=0$, we see from (\ref{tconstEOM})
that a surface ``hanging down'' at generic $\theta=const$ is not
extremal: although the left hand side of (\ref{tconstEOM}) vanishes,
the right hand side does not.
However we are most interested in the limit where the subregion is
maximal, \ie\ when the cap-like region becomes the entire hemisphere
on the $S^{d-1}$: this is when $\theta={\pi\over 2}$. It can be seen
that this now is a solution to the extremization equation above: 
the right hand side vanishes with $\cos{\pi\over 2}=0$. For these
limiting surfaces, the area becomes
\be\label{thetaConstSurf}
\theta={\pi\over 2} :\qquad\qquad
       {S\over 4G_{d+1}} =\ {2l^{d-1} V_{S^{d-2}}\over 4G_{d+1}} \int_\epsilon^1
       {d\tau\over\tau^{d-1}} {1\over\sqrt{1-\tau^2}}\ \
       \xrightarrow{\ dS_4\ }\ \ {\pi l^2\over G_4} {1\over\epsilon}\ .
\ee
Thus this again recovers de Sitter entropy as the coefficient of the
area law divergence as in the previous case (\ref{EEdSent}),
(\ref{EEdS4ent}). The geometry of these surfaces away from precisely
$\theta'=0$ is however somewhat different.
This can be seen in some detail in the $dS_3$ case, \ie\ $d=2$. Since
$\theta$ is now a cyclic coordinate in (\ref{areaFnTheta}), we have a
conserved quantity $A={\del L\over\del{\theta'}}$ giving
\be
\theta' = {A\tau\over \sqrt{(1-\tau^2)(1+A^2\tau^2)}}\ .
%S = l \int_\epsilon^{\tau_*} {d\tau\over\tau}
%{1\over \sqrt{(1-\tau^2)(1+A^2\tau^2)}}\ ,
\ee
For $A\ra\infty$, these are null surfaces
\be\label{thetaNullsurf}
\theta'= 1/\sqrt{1-\tau^2}\quad \Rightarrow\quad
\tau = \sin(\theta_0-\theta)\ ; \quad
\tau_{max}=\sin\theta_0\ \ra\ 1\ \ {\rm as}\ \ \theta_0\ra \pi/2\ .
\ee
These are disconnected surfaces with (minimal) vanishing area.
For precisely $A=0$, these are $\theta=const$ surfaces which ``hang down''
into the bulk without turning, but slightly different geometrically
from the ones in the equatorial plane earlier: to see this, consider
small $A=\varepsilon$ and $\theta_0={\pi\over 2}$.
Then as $\tau\ra 1$, we have
\be\label{dS3dimple}
\theta'\sim \pm A\tau/\sqrt{1-\tau^2}
%\xrightarrow{\tau\ra 1}\ \pm {A\over \sqrt{2(1-\tau)}}\ ,
\qquad \Rightarrow \qquad \theta - \theta_0 \sim \pm  A\sqrt{2(1-\tau)}\ .
\ee
We see that for any nonzero infinitesimal $A$, the surface at
$\theta\sim\theta_0$ has a tendency to turn as $\tau\ra 1-O(A^2)$ with
$\theta'\ra\infty$. However $\theta$ is essentially constant till very
near $\tau=1$ and then $\theta-\theta_0\sim O(A)\sim O(\varepsilon)$. Thus 
these surfaces at $\theta={\pi\over 2}$ stretch from $I^+$ to $\tau\sim 1$
where they acquire an $O(A)$ ``dimple'' where $\theta'\ra\infty$. The
full connected surface (after joining an equivalent surface from $I^-$
in $P$) thus has an $O(A)$ constriction at the neck in
Figure~\ref{dSstSurfEnt2}, and so is not smooth unlike the $\theta'=0$
surface at $\theta={\pi\over 2}$\ (or the red curves in
Figure~\ref{dSstSurfEnt}).

For other dimensions, it appears difficult to identify exact solutions
although they may well exist. Null surfaces of course continue to
arise as in (\ref{thetaNullsurf}). Considering 
$\theta'=0$ surfaces, let us now consider the neighbourhood of the
$\theta={\pi\over 2}$ extremal surface: parametrizing this as\
$\theta(\tau)={\pi\over 2}-\delta\theta(\tau)$\ to $O(\delta\theta)$
gives $\cos\theta\sim \delta\theta(\tau)$ and the linearized equation 
\be
   {d\over d\tau} \left({-\sqrt{1-\tau^2}\over \tau^{d-1}}\
   {d\delta\theta(\tau)\over d\tau}\right) =
   {d-2\over \tau^{d-1}} {\delta\theta(\tau)\over\sqrt{1-\tau^2}}\ .
\ee
The solution that is regular as $\tau\ra 0,1$, and in addition exhibits
$\theta'$ monotonically increasing till $\theta'\ra\infty$ as $\tau\ra 1$
is\  $\delta\theta(\tau)=\tau^d~{}_2F_1\big({1+d-\sqrt{d^2+2d-7}\over 4} ,
{1+d+\sqrt{d^2+2d-7}\over 4} ,1+{d\over 2} ;\ \tau^2\big),$\
involving the hypergeometric function ${}_2F_1$.\ 
Since\ $\delta\theta(\tau)$ encodes the infinitesimal linearization about
$\theta={\pi\over 2}$~, regularity of this solution implies that the
surface $\theta(\tau)$ has near-constant $\theta\sim {\pi\over 2}$ and
acquires an infinitesimal ``dimple'' at $\tau\sim 1$ similar to the
$O(A)$ dimple in the $dS_3$ surface (\ref{dS3dimple}). It then joins
an equivalent surface in the past universe. The full connected surface
stretching from $I^+$ to $I^-$ thus has a constriction at the neck 
(Figure~\ref{dSstSurfEnt2}) and is not smooth. In the absence of the
detailed solution, it is difficult to check if the $\theta'=0$ 
surface at $\theta={\pi\over 2}$ has minimal area although the area
integral (\ref{thetaConstSurf}) is identical to (\ref{EEdSent}),
(\ref{EEdS4ent}).

\section{Discussion}

We have seen that 4-dim de Sitter entropy ${l^2V_{S^2}\over 4G_4}={\pi
  l^2\over G_4}$ which is the area of the cosmological event horizon
for regions $N$ and $S$ in the static patch coordinatization
(Figure~\ref{dSstSurfEnt}) arises as the coefficient of the divergent
area ${\pi l^2\over G_4} {1\over\epsilon}$ of certain codim-2 real
timelike extremal surfaces.  These wrap an $S^1$ and stretch in the
bulk time direction from the future boundary $I^+$ in $F$ to the past
boundary $I^-$ in $P$, the areas along the $S^1$ and the time
direction compensating for $V_{S^2}$. These surfaces all lie in a
boundary Euclidean time slice of the bulk space, either (i) in some
equatorial plane of the $S^2$, where they exclude the regions behind
the horizons, or (ii) on the $w=const$ slice. As the boundary
subregion approaches all of $I^\pm$, they all pass through the
bifurcation region (with only the divergent term): the ones in the
equatorial planes can be seen to have minimal area. The vicinity of
the bifurcation region behaves a bit like $AdS$ with regard to area
extremization.  These surfaces are in some sense rotated versions of
surfaces of Hartman and Maldacena \cite{Hartman:2013qma} in the $AdS$
black hole (which itself after a rotation resembles the present de
Sitter static coordinatization).

The restriction to a boundary Euclidean time slice which encodes a
symmetry direction gives codim-2 surfaces, consistent with the
${l^2\over G_4}$ scaling of de Sitter entropy. The fact that the
divergence coefficient arises independent of which particular boundary
Euclidean time slice is used suggests that there exists some
formulation which makes manifest this independence on the particular
slice. The boundary Euclidean time slice is of course reminiscent of
the constant time slice containing spatial subsystems in the usual
formulation of entanglement entropy, which we will elaborate on below.

In the context of $dS/CFT$
\cite{Strominger:2001pn,Witten:2001kn,Maldacena:2002vr}, de Sitter
space is conjectured to be dual to a hypothetical Euclidean
non-unitary CFT that lives on the future boundary ${\cal I}^+$, with
the dictionary $\Psi_{dS}=Z_{CFT}$\ \cite{Maldacena:2002vr}, where
$\Psi_{dS}$ is the late-time Hartle-Hawking wavefunction of the
universe with appropriate boundary conditions and $Z_{CFT}$ the dual
CFT partition function. The dual CFT$_d$ energy-momentum tensor
correlators reveal central charge coefficients ${\cal C}_d\sim
i^{1-d}{l^{d-1}\over G_{d+1}}$ in $dS_{d+1}$ (effectively analytic
continuations from $AdS/CFT$). This is real and negative in $dS_4$ so
that $dS_4/CFT_3$ is reminiscent of ghost-like non-unitary
theories. In \cite{Anninos:2011ui}, a higher spin $dS_4$ duality was
conjectured involving a 3-dim CFT of anti-commuting $Sp(N)$ (ghost)
scalars, studied previously in
\cite{LeClair:2006kb,LeClair:2007iy}\ (see also
\eg\ \cite{Bousso:2001mw,Balasubramanian:2002zh,
  Harlow:2011ke,Ng:2012xp,Das:2012dt,Anninos:2012ft}).

As we have seen, the areas of the codim-2 extremal surfaces here scale
as ${l^{d-1}\over G_{d+1}}$ but have a different numerical factor in
other dimensions: perhaps this is not unexpected since $dS/CFT$ away
from $dS_4$ appears more exotic.  Relatedly in $dS_4$, interpreting
the extremal surface area as an area law divergence in a dual CFT, one
might worry about the detailed significance of the coefficient:
rescaling the ultraviolet cutoff changes the precise coefficient. The
bulk surface stretches from the future boundary $I^+$ regulated at
$\epsilon={\epsilon_c\over l}$ (expressed in terms of the de Sitter
scale $l$), and passes through the bifurcation region at $\tau=1$
(which contains the turning point) and its area is unambiguous
however. Perhaps it is noteworthy that in the limit of the subregion
being the full space, the surface in question almost cloaks the
horizons suggesting that the area in some sense encodes degrees of
freedom behind the horizons (although these are Cauchy horizons for
the future/past universes).

In the static patch coordinatization here, since the boundary at
$I^\pm$ is Euclidean $R_w\times S^{d-1}$, it is reasonable to imagine
the dual to de Sitter space to comprise two copies of the dual
Euclidean nonunitary CFT on a cylindrical Euclidean space of the form
$R_w\times S^{d-1}$. Noting de Sitter entropy, one might be tempted to
regard it as an ``entangled state'' of $CFT_F\times CFT_P$ \ie\ two
copies of the dual CFT, with de Sitter entropy appearing as the
coefficient of some ``generalized entanglement entropy''. This cannot
be entanglement in the usual sense since the dual CFTs are Euclidean:
however the presence of a translation symmetry along some boundary
direction taken as Euclidean time allows formulating a generalized
entanglement in a formal manner along the usual lines. The connected
extremal surfaces here (Figure~\ref{dSstSurfEnt}) stretching from
$CFT_F$ at $I^+$ to $CFT_P$ at $I^-$ appear to corroborate this
interpretation\footnote{From \cite{Maldacena:2002vr}, the $CFT$
  partition function $Z_{CFT}=\Psi_{dS}\sim e^{iI_{dS}}$ has
  imaginary pieces in a semiclassical expansion: however bulk
  expectation values involve the probability
  $|\Psi_{dS}|^2=\Psi_{dS}^*\Psi_{dS}$ where these cancel, which
  also might be taken to point to $CFT_F\times CFT_P$ for bulk
  physics.}. The negative central charge of $dS_4/CFT_3$ suggests that
$CFT_{F,P}$ are akin to ghost-CFTs, as mentioned above.

With a view to elaborating further, we first recall that
certain complex codim-2 extremal surfaces were found to give negative
areas in $dS_4$ \cite{Narayan:2015vda,Narayan:2015oka}, consistent
with the negative central charge\ (amounting to analytic continuation
from Ryu-Takayanagi in $AdS$). Towards gaining some insight on
generalizations of entanglement entropy to ghost-like theories and
negative entanglement, certain investigations were carried out in
\cite{Narayan:2016xwq} in toy 2-dim ghost-CFTs using the replica
formulation (giving $S<0$ for $c<0$ ghost-CFTs under certain conditions)
and in quantum mechanical toy models of ``ghost-spins'' (reviewed
briefly in Appendix~\ref{sec:gs}) via reduced density matrices.
A single ghost-spin is defined as a 2-state spin variable
with indefinite inner product\ $\lan\ua|\ua\ran = 0 = \lan\da|\da\ran$
and $\lan\ua|\da\ran = 1 = \lan\da|\ua\ran$,\ akin to the inner
products in the $bc$-ghost system\ (in contrast, a single spin
has\ $\lan\ua|\ua\ran = 1 = \lan\da|\da\ran$). Then the states
$|\pm\ran={1\over\sqrt{2}}(|\ua\ran\pm |\da\ran)$ satisfy
$\lan\pm |\pm\ran=\pm 1$. A two ghost-spin state then has norm
\be\label{norm2gs}
|\psi\ran=\psi^{\al\beta}|\al\beta\ran:\qquad
\lan\psi|\psi\ran=\gamma_{\alpha\kappa} \gamma_{\beta\lambda}
\psi^{\alpha\beta} {\psi^{\kappa\lambda}}^*\ ,\qquad \gamma_{++}=1,\ \ 
\gamma_{--}=\lan -|-\ran=-1\ ,
\ee
where $\gamma_{\al\beta}$ is the indefinite metric. Thus although states
$|-\ran$ have negative norm, the state $|-\ran|-\ran$ has positive
norm. Ensembles of ghost-spins were developed further in
\cite{Jatkar:2016lzq} for entanglement properties. In
\cite{Jatkar:2017jwz} certain 1-dim ghost-spin chains with specific
nearest-neighbour interactions were found to yield $bc$-ghost CFTs in
the continuum limit. %, \ie\ these ghost-spin chains are in the same universality class as those ghost-CFTs.
Some ongoing work deals with
certain $N$-level generalizations of the 2-level ghost-spins above. In
this light, one might regard appropriate 3-dimensional $N$-level
ghost-spin chains as approximating the ghost-$CFT_3$s above and
thereby possibly $dS_4$ with ${l^2\over G_4}\sim N$\ (the negative
central charge suggests $dS_4$ is like a ghost-CFT more generally, not
just in the higher spin context of \cite{Anninos:2011ui}). Thinking
thereby of appropriate 3-dim ghost-spin systems as microscopic
realizations in the same universality class as ghost-CFTs dual to
$dS_4$, we can study entanglement properties for states in the
ghost-spin system.
Then consider an entangled state
of the form
\be\label{dsEntstate}
|\psi\rangle = \sum \psi^{i_n^F,i_n^P} |i_n^F\ran |i_n^P\ran
\ee
where $\psi^{i_n^F,i_n^P}$ are coefficients entangling a generic ghost-spin
configuration $|i_n^F\ran$ from $CFT_F$ at $I^+$ with an identical one
$|i_n^P\ran$ from $CFT_P$ at $I^-$.  (Entangled states of this schematic
form appear in \cite{Bousso:2001mw}: the state (\ref{dsEntstate})
however involves entanglement between states of ghost-spin systems,
or ghost-CFTs.) The state (\ref{dsEntstate}) is akin to a correlated
ghost-spin state with an even number of ghost-spins, as discussed in
\cite{Jatkar:2016lzq,Jatkar:2017jwz}.
It necessarily has positive norm,
\be\label{dsEntstateNorm}
\sum_{i_n^F,i_n^P} \gamma_{i_n^F,i_n^F}\ \gamma_{i_n^P,i_n^P}\ \psi^{i_n^F,i_n^P}
\big(\psi^{i_n^F,i_n^P}\big)^*
\ \ \xrightarrow{\ F\equiv P\ }\ \ \sum_{i_n} |\psi^{i_n^F,i_n^P}|^2
\ee
using the form (\ref{norm2gs}), since we are entangling identical
states $i_n^F$ and $i_n^P$: thus it has positive entanglement, as in
\cite{Jatkar:2016lzq,Jatkar:2017jwz}. Since each constituent state
$|i_n^{F,P}\ran$ is $N$-level, \ie\ with $N$ internal degrees of
freedom, the entanglement entropy scales as\ $N\sim {l^2\over G_4}$.
The state (\ref{dsEntstate}) is akin to the thermofield double dual to
the eternal $AdS$ black hole \cite{Maldacena:2001kr}. This suggests
the speculation that 4-dim de Sitter space is perhaps approximately
dual to $CFT_F\times CFT_P$ in the entangled state (\ref{dsEntstate})
and the generalized entanglement entropy of the latter scales as de
Sitter entropy. It is also interesting to speculate about ER=EPR
\cite{Maldacena:2013xja} and so on in this context\footnote{I thank
  Juan Maldacena for an interesting correspondence in this regard.}:
the roles of bulk time evolution and boundary Euclidean time evolution
are likely to make this structure somewhat different from $AdS$. In
particular, the bulk time evolution operator, which maps states at
$I^-$ to $I^+$\ \cite{Witten:2001kn}, can be used to argue that the
states (\ref{dsEntstate}) are unitarily equivalent to maximally
entangled states
$|\psi\rangle = \sum \psi^{i_n^F,i_n^F} |i_n^F\ran |i_n^F\ran$ in two
$CFT_F$ copies solely at the future boundary. We hope to explore these
further.

\vspace{7mm}

{\footnotesize \noindent {\bf Acknowledgements:}\ \ It is a pleasure
  to thank Dileep Jatkar for interesting discussions and initial
  collaboration on this work. I also thank Sumit Das and Sandip
  Trivedi for interesting conversations over the last two years, Andy
  Strominger for interesting conversations at SpentaFest, ICTS, and
  Amitabh Virmani for useful comments on a draft. This work is
  partially supported by a grant to CMI from the Infosys Foundation.
}

\vspace{-2mm}

\appendix

\section{Reviewing extremal surfaces in Poincare $dS$}\label{sec:extSurPoinc}

Reviewing \cite{Narayan:2015vda,Narayan:2015oka}, consider de Sitter
space in Poincare slicing,\
$ds^2 = {R_{dS}^2\over\tau^2} (-d\tau^2 + dw^2 + dx_i^2)$.\ One of
the $d$ spatial directions $w$ is regarded as boundary Euclidean time.
On a $w=const$ slice, consider a strip subregion at $I^+$
with width $\Delta x=l$ and extremal surfaces anchored at the
subregion boundary and dipping into the bulk towards the past. The
area functional is
\be
S_{dS} = {1\over 4G_{d+1}} 
\int \prod_{i=1}^{d-2} {R_{dS}dy_i\over\tau} {R_{dS}\over\tau} 
\sqrt{d\tau^2 - dx^2} = {R_{dS}^{d-1} V_{d-2}\over 4G_{d+1}} 
\int {d\tau\over\tau^{d-1}} \sqrt{1- {\dot x}^2}\ ,
\ee
the sign under the square root reflecting real timelike surfaces.
The extremization gives
\be\label{RTdS02}
\Big({dx\over d\tau}\Big)^2 \equiv {\dot x}^2
= {B^2\tau^{2d-2}\over 1+B^2\tau^{2d-2}}\ ,\qquad
S_{dS} = {R_{dS}^{d-1}\over 4G_{d+1}} V_{d-2} \int_\epsilon^{\tau_0}
{d\tau\over\tau^{d-1}} {1\over\sqrt{1+B^2\tau^{2d-2}}}\ .
\ee
We see that the square root in the area integral is quite different
from (\ref{HMsurf}): there is no minus sign so the behaviour is
monotonic with $B$ (no turning point exists where ${\dot x}^2\ra\infty$
for real $\tau$; complex extremal surfaces with turning points exist 
along imaginary time paths $\tau=iT$, amounting to analytic continuation
from Ryu-Takayanagi in $AdS$). The area is
maximised for $B=0$, giving maximally
timelike surfaces (${\dot x}=0$) which simply ``hang'' down into the
bulk. For surfaces stretching all the way to $\tau=-\infty$, the area
becomes
\be
S_{dS} = {R_{dS}^{d-1}\over 4G_{d+1}} V_{d-2} \int_{-\infty}^{-\epsilon}
{d\tau\over\tau^{d-1}}\ \ \longrightarrow\ \
{R_{dS}^{2}\over 4G_{4}} {V_{1}\over\epsilon}\quad [dS_4]\ ,
\ee
independent of the size of the subregion. There are no finite
cutoff-independent pieces for these surfaces since those contributions
die at $|\tau|\ra\infty$.

Now consider spherical subregions on ${\cal I}^+$, with
radius $l$ satisfying $0\leq r\leq l$. Towards studying spherical
entangling surfaces, we parametrize the spatial part of the $dS$
metric in polar coordinates. The $w=const$ slice has metric\
$ds^2 = {R_{dS}^2\over\tau^2} (-d\tau^2 + dr^2 + r^2 d\Omega_{d-2}^2)$.\
The bulk codim-2 surface $r=r(\tau)$ has area functional in Planck units
\be\label{RTdS01}
S_{dS} = {1\over 4G_{d+1}} 
\int \prod_{i=1}^{d-2} {R_{dS} r d\Omega_i\over\tau} {R_{dS}\over\tau} 
\sqrt{d\tau^2-dr^2} = {R_{dS}^{d-1} \Omega_{d-2}\over 4G_{d+1}} 
\int {d\tau\over\tau^{d-1}} r^{d-2} \sqrt{1-{\dot r}^2}\ .
\ee
The variational equation of motion for an extremum\ 
${\del\over\del\tau} ({\del {\cal L}\over\del {\dot r}}) = 
{\del {\cal L}\over \del r}$\ leads to\ (with
${dr\over d\tau}\equiv {\dot r}$)\ 
\be\label{EOM-extsurf}
{\del\over\del\tau} \Big( {r^{d-2}\over\tau^{d-1}} 
{{\dot r}\over\sqrt{1-{\dot r}^2}} \Big) = {d-2\over \tau^{d-1}} 
r^{d-3} \sqrt{1-{\dot r}^2} \qquad \longrightarrow\qquad
%\ee It can be seen that \be\label{sphExtSur}
r(\tau) = \sqrt{l^2 + \tau^2}\ .
\ee
This satisfies the boundary conditions\ $r\ra l$ as $\tau\ra 0$.
Unlike the strip case, there are no parameters here for $d>2$.
For $\tau$ real, there is no bulk turning point where ${\dot r}\ra\infty$:
instead $r^2\ra\tau^2$ asymptotically. Since $r(\tau)\geq l$, this
surface bends ``outwards'' from the subregion boundary (all interior
points within the subregion satisfy $0\leq r\leq l$). This real
timelike surface ${\dot r} \leq 1$ does not ``end'' at any finite
$\tau$: considering the whole $\tau$-range, we obtain\ 
\be
S_{dS}={R_{dS}^{d-1} \Omega_{d-2} l \over 4G_{d+1}} 
\int_{-\infty}^{-\epsilon} {d\tau\over\tau^{d-1}} (l^2+\tau^2)^{(d-3)/2}\quad
\xrightarrow{\ dS_4\ }\quad  {R_{dS}^2\over 4G_4} {A_1\over\epsilon}\ ,
\ee
with $\tau_{UV}=-\epsilon$, and $A_1=2\pi l$ the $S^1$ interface
area ($dS_4$). For $dS_5$, we obtain\
$S_{dS} = {R_{dS}^3\over 8G_5} {A_2\over\epsilon^2} + 
{\pi R_{dS}^3\over 2G_5} \log {l\over\epsilon}$, with $A_2=4\pi l^2$
the $S^2$ interface area. Again the finite pieces vanish.

\section{Reviewing ghost-spins and ghost-spin chains}\label{sec:gs}

Here we review ghost-spins and ghost-spin chains, based on
\cite{Narayan:2016xwq,Jatkar:2016lzq,Jatkar:2017jwz}. 
A single ghost-spin is defined as a 2-state spin variable
with indefinite inner product\ $\lan\ua|\ua\ran = 0 = \lan\da|\da\ran$
and $\lan\ua|\da\ran = 1 = \lan\da|\ua\ran$.\
Then $|\pm\ran={1\over\sqrt{2}}(|\ua\ran\pm |\da\ran)$ have
positive/negative norm.
We normalize positive/negative norm states with norm $\pm 1$
respectively. Now consider the two ghost-spin state (\ref{norm2gs}).
The full density matrix is
$\rho=|\psi\rangle\langle\psi| = \sum \psi^{\alpha\beta} {\psi^{\kappa\lambda}}^* 
|\alpha\beta \rangle\langle \kappa\lambda|$.
Tracing over one of the ghost-spins leads to a reduced density matrix
$(\rho_A)^{\alpha\kappa} = \gamma_{\beta\lambda} \psi^{\alpha\beta} {\psi^{\kappa\lambda}}^* = \gamma_{\beta\beta}  \psi^{\alpha\beta} {\psi^{\kappa\beta}}^*$,
\bea\label{rhoA2gs}
\qquad\qquad 
(\rho_A)^{++} =\ |\psi^{++}|^2 - |\psi^{+-}|^2\ , &\quad&
(\rho_A)^{+-} =\ \psi^{++} {\psi^{-+}}^* - \psi^{+-} {\psi^{--}}^*\ , 
\nonumber\\
(\rho_A)^{-+} =\ \psi^{-+} {\psi^{++}}^* - \psi^{--} {\psi^{+-}}^*\ , &\quad& 
(\rho_A)^{--} =\ |\psi^{-+}|^2 - |\psi^{--}|^2\ , 
\eea
for the remaining ghost-spin.
Then $tr \rho_A = \gamma_{\alpha\kappa} (\rho_A)^{\alpha\kappa} =
(\rho_A)^{++} - (\rho_A)^{--}$. 
Thus the reduced density matrix is normalized to have\ 
$tr \rho_A = tr \rho = \pm 1$ depending on whether the state 
(\ref{norm2gs}) is positive or negative norm.
The entanglement entropy calculated as the von Neumann entropy of 
$\rho_A$ is\ $S_A = -\gamma_{\alpha\beta} (\rho_A \log \rho_A)^{\alpha\beta}$,
perhaps best defined in terms of a mixed-index reduced density matrix
$(\rho_A)^\alpha{_\kappa}$.
As an illustration, consider a simple family of states
\cite{Narayan:2016xwq} with a diagonal RDM: setting
${\psi^{-+}}^* = \psi^{+-} {\psi^{--}}^*/\psi^{++}$\ in the states
(\ref{norm2gs}) gives\
%$\langle\psi| \psi\rangle = (|\psi^{++}|^2 - |\psi^{+-}|^2) 
%(1 + |\psi^{--}|^2/|\psi^{++}|^2 ) = \pm 1$, and (\ref{rhoA2gs}) gives
\bea\label{Ex:2gsEE}
&& (\rho_A)^{\al\beta}|\al\ran\lan\beta| = \pm x |+ \rangle\langle +|\
\mp\ (1-x) |- \rangle\langle -|\ , \qquad
x = {|\psi^{++}|^2\over |\psi^{++}|^2 + |\psi^{--}|^2} \qquad [0 < x < 1] ,
\nonumber\\
&& (\rho_A)_\alpha^\kappa = \gamma_{\alpha\beta} (\rho_A)^{\beta\kappa}:
\qquad (\rho_A)^+_+ = \pm x ,\qquad (\rho_A)^-_- = \pm (1-x) ,\quad
\eea
where the $\pm$ pertain to positive/negative norm states respectively\
(note that\ $tr\rho_A = (\rho_A)^+_+ + (\rho_A)^-_- = \pm 1$).
The location of the negative eigenvalue is different for
positive/negative norm states, leading to different results for the
von Neumann entropy. Now $\log\rho_A$ simplifies to\  
$(\log\rho_A)^+_+ = \log (\pm x)$ and $(\log\rho_A)^-_- = \log (\pm (1-x))$.
The entanglement entropy
$S_A = -\gamma_{\alpha\beta}(\rho_A\log\rho_A)^{\alpha\beta}$ becomes\
$S_A = - (\rho_A)^+_+ (\log\rho_A)^+_+ - (\rho_A)^-_- (\log\rho_A)^-_-$\ 
and so
\be\label{Ex:2gsEE2}
\langle\psi| \psi\rangle \gtrless 0:\qquad
S_A = - (\pm x)\log (\pm x) - (\pm (1-x)) \log (\pm (1-x)) \ .
%\langle\psi| \psi\rangle < 0:\quad S_A =  x\log (-x) + (1-x) \log (-(1-x))\ .
\ee
For positive norm states, $S_A$ is manifestly positive since $x<1$,
just as in an ordinary 2-spin system.\ Negative norm states give a 
negative real part for EE ($x<1$ and the logarithms are negative), and
an imaginary part (the simplest branch has $\log (-1)=i\pi$).\

Note that restricting to the subspace\
$|\psi\ran=\psi^{++}|++\ran+\psi^{--}|--\ran$, we obtain solely
positive norm states and positive RDM and entanglement. For ensembles
with an even number of ghost-spins, such correlated ghost-spin states
always exist comprising positive norm subsectors. This leads to
(\ref{dsEntstate}), (\ref{dsEntstateNorm}). Odd ghost-spins behave
differently: \eg\
$|\psi\ran=\psi^{++\ldots}|++\ldots\ran+\psi^{--\ldots}|--\ldots\ran$
has norm $\lan\psi|\psi\ran=|\psi^{++\ldots}|^2 + (-1)^n|\psi^{--\ldots}|^2$
and mixed-index RDM components\
$(\rho_A)^+_+=|\psi^{++\ldots}|^2,\ (\rho_A)^-_-=(-1)^n|\psi^{--\ldots}|^2$.
This is not positive definite for $n$ odd (even if $\lan\psi|\psi\ran>0$).
Ensembles of ghost-spins and spins also show interesting
entanglement patterns \cite{Jatkar:2016lzq}.

It is natural to expect that infinite 1-dim chains of ghost-spins lead
in a continuum limit to 2-dim ghost-CFTs, akin to the well-known fact
that the Ising spin chain at criticality is described by a CFT of free
massless fermions. Consider spin variables $\sigma_{bn},\ \sigma_{cn}$
satisfying  %the (anti-)commutation relations
\be\label{sigmaCommRelns}
\{ \sigma_{bn},\ \sigma_{cn} \} = 1\ ,\qquad [\sigma_{bn}, \sigma_{bn'}] =
[\sigma_{cn}, \sigma_{cn'}] = [\sigma_{bn}, \sigma_{cn'}] = 0\ ,
\ee
which are consistent with the off-diagonal inner product between
ghost-spin states.  These spin variables are self-adjoint and act on
two states $|\ua\ran,\ |\da\ran$, at each lattice site $n$, as 
\be
\sigma_{bn}^\dag=\sigma_{bn} ,\ \ 
\sigma_{cn}^\dag=\sigma_{cn}\ ;\qquad \sigma_b|\da\ran = 0 ,\quad
\sigma_b|\ua\ran = |\da\ran ,\quad \sigma_c|\ua\ran = 0 ,\quad
\sigma_c|\da\ran = |\ua\ran .  
\ee
Consider now a 1-dim ghost-spin chain with nearest neighbour
interaction Hamiltonian
\be\label{Hgssigmabc}
H = J \sum_n \left( \sigma_{b(n)} \sigma_{c(n+1)} 
+ \sigma_{b(n)}\sigma_{c(n-1)} \right)\ .
\ee
This is not quite Ising-like: in fact it describes a ``hopping'' type
Hamiltonian, which kills an $\ua$-spin at site $n$ and creates it at
site $n\pm 1$, so that $\ua_n$ hops to $\ua_{n\pm 1}$.

The $\sigma_{bn}, \sigma_{cn}$ above are analogous to the $b_n,c_n$
operators of the $bc$-ghost CFT, satisfying $\{b_n, c_m\} = \delta_{n+m,0}$:
however $\sigma_{bn},\ \sigma_{cn}$ are bosonic, commuting at distinct
lattice sites. Fermionic ghost-spin variables constructed via a
Jordan-Wigner transformation
\be\label{JW}
a_{bn} = \prod_{k=1}^{n-1} i(1-2\sigma_{ck}\sigma_{bk}) \sigma_{bn}\ ,\qquad
a_{cn} = \prod_{k=1}^{n-1} (-i)(1-2\sigma_{ck}\sigma_{bk}) \sigma_{cn}\ ,
\ee
\be\label{fermgsops-abc}
{\rm give}\qquad\qquad\qquad\quad
\{ a_{bi}, a_{cj} \} = \delta_{ij}\ ,\qquad \{ a_{bi}, a_{bj} \} = 0\ , \qquad
\{ a_{ci}, a_{cj} \} = 0\ .\qquad\qquad
\ee
Unlike the $\sigma$ spin operators, these anticommute not just at
the same site $i$ but also at distinct sites $i, j$.\
The states satisfy\
$a_b|\da\ran = 0 ,\ a_b|\ua\ran = |\da\ran ,\
a_c|\ua\ran = 0 ,\ a_c|\da\ran = |\ua\ran ,\
\lan \da| a_b = 0 ,\ \lan \ua| a_b = \lan\da| ,\
\lan \da| a_c = \lan \ua| ,\ \lan \ua| a_c = 0 .$
To construct states and their inner products, we have to be careful
about the ordering of the operators and the spin excitations. We adopt
the convention that 
%\be\label{fgsOrdering}
$\lan\ua\ua| \da\da\ran = 1 ,\
|\ura{\ua\ua}\ran = a_{c1} a_{c2} |\da\da\ran ,\ 
\lan\ula{\da\da}| = \lan \ua\ua| a_{b2} a_{b1}$, giving\
$\lan\ula{\da\da} |\ura{\ua\ua}\ran = 1$,
%\ee
illustrating two fermionic ghost-spins.
The underlining right arrow in the ket displays the order of the
operator excitations to be increasing to the right; the underlining
left arrow in the bra shows the order as increasing to the left.
Then a state $\psi_1|\da\da\ran + \psi_2 |\ura{\ua\ua}\ran$\ \ gives\ \
$\psi_1^*\psi_2 \lan\ua\ua| a_{b_2} a_{b1}\ a_{c1} a_{c2} |\da\da\ran 
+ \psi_2^*\psi_1 \lan\ua\ua |\da\da\ran = \psi_1^*\psi_2 + \psi_2^*\psi_1$,\ \
the expected indefinite norm.

In terms of the fermionic ghost-spin variables, the Hamiltonian
(\ref{Hgssigmabc}) becomes
\be
H\ \ra\ \ i J a_{bn} ( a_{c(n+1)} - a_{c(n-1)} )\ \ \sim\ \ -b\del c\ ,
\ee
which is the lattice discretization of the $bc$-ghost CFT. Using
momentum space variables\
$b_k = {1\over\sqrt{N}} \sum_n e^{ikn} a_{bn} ,\ 
c_k = {1\over\sqrt{N}} \sum_n e^{ikn} a_{cn}$,\ we obtain\
$\{ b_k, c_{k'} \} = \delta_{k+k',0},\ \{ b_k, b_{k'} \} = 0 =
\{ c_k, c_{k'} \}$.\
Reinstating the lattice spacing $a$, we obtain in the continuum limit
$a\ra 0$
\be
H = 2J \sum_k \sin (k'a)\ b_k c_{k'}\ \delta_{k+k',0} =
2Ja \sum_k k b_{-k}c_k \xrightarrow{ J\sim 1/2a } 
  \sum_{k>0} k \left( b_{-k}c_k + c_{-k}b_k \right) + \zeta\ ,
\ee
with $\zeta$ the normal ordering constant giving the zero point energy.
The scaling $J\sim {1\over 2a}$ ensures that the nearest neighbour
lattice interaction leads to a nontrivial continuum interaction.\ 
Consider now the symmetries of the ghost-spin 
chain Hamiltonian (\ref{Hgssigmabc}): first, the phase rotation symmetry\
$\sigma_{b(n)}\ra e^{i\al} \sigma_{b(n)},\
\sigma_{c(n+1)} \ra e^{-i\al} \sigma_{c(n+1)}$,\
is the microscopic reflection of the $U(1)$ symmetry in the continuum 
$bc$-CFT. Also
\be\label{gschainScalingsymm}
a\ra \xi^{-1} a\ ,\quad H\ra \xi H\ ,\qquad 
\sigma_{b(n)}\ra \xi^\lambda \sigma_{b(n)}\ ,\qquad
\sigma_{c(n+1)} \ra \xi^{1-\lambda} \sigma_{c(n+1)}\ ,
\ee
is a global scaling symmetry of the ghost-spin variables
$(\sigma_b,\sigma_c)$ for any constant $\lambda$: this underlies 
the conformal symmetry of the $bc$-CFT, with conformal weights
$(h_b, h_c) = (\lambda,1-\lambda)$. Further details appear in 
\cite{Jatkar:2017jwz}.\ This suggests that ghost-spins are microscopic
building blocks of ghost-like CFTs in general, including 3-dim ones.

%\vspace{6mm}
%\newpage

\end{document}